\documentclass[journal]{IEEEtran}
\ifCLASSINFOpdf
\else
   \usepackage[dvips]{graphicx}
\fi
\usepackage{url}

\usepackage{graphicx}

\usepackage{amsmath,amssymb,threeparttable,placeins,makecell,stfloats,cite}
\usepackage[usenames]{color}
\usepackage{amsfonts}
\usepackage{latexsym}
\usepackage{lipsum}
\usepackage{floatrow}
\usepackage{cuted}
\usepackage{lipsum,booktabs}
\usepackage{subcaption}

\newtheorem{theorem}{{{\textit{Theorem}}}}

\newtheorem{lemma}{{{\textit{Lemma}}}}

\newtheorem{definition}{{{\textit{Definition}}}}

\newtheorem{remark}{{{\textit{Remark}}}}

\newtheorem{example}{{{\textit{Example}}}}

\newcounter{cases}
\newcounter{subcases}[cases]

\begin{document}
\title{Construction of Optimal Binary Z-Complementary Code Sets with New Lengths}
\author{Gobinda~Ghosh,~
        Sudhan~Majhi,~
        Shubabrata Paul
 {
}

}

\IEEEpeerreviewmaketitle
\maketitle
\begin{abstract}
Z-complementary code sets (ZCCSs) are used in multicarrier code-division multiple access (MC-CDMA) systems, for interference-free communication over multiuser and quasi-asynchronous environments. 
 In this letter, we  propose three new constructions of optimal binary $\left(R2^{k+1},2^{k+1}, R\gamma,\gamma\right)$-ZCCS, $\left(R2^{k+1},2^{k+1}, R2^{m_{2}},2^{m_{2}}\right)$-ZCCS and $\left(2^{k+1},2^{k+1},3\gamma,2\gamma\right)$-ZCCS
  based on generalized Boolean functions (GBFs), where $\gamma=2^{m_{1}-1}+2^{m_{1}-3}, m_{1}\geq 5, k\geq 1,m_{2}\geq 1$ and $R$ is any even number. The proposed ZCCSs cover many unreported lengths and large set sizes.

\end{abstract}
\section{Introduction}
\IEEEPARstart{M}{ulticarrier} code-division multiple access (MC-CDMA) is a popular wireless technique used in many communication systems due to its efficient fast Fourier transform (FFT) based implementation, resistance to intersymbol interference, and high spectral efficiency \cite{chen2007next}. However, MC-CDMA can be affected by multiple-access interference (MAI)\cite{carey2003comparison} and multipath interference (MPI) \cite{nagaradjane2009multipath}. Adopting proper spreading codes, such as complete complementary codes (CCC) \cite{rathinakumar2008complete} and Z-complementary code sets (ZCCSs) \cite{Davis}, aforementioned problems can be resolved. ZCCS has large code size thus support larger number of users compared to CCC. However, there is a limitation of having flexibility on length and size of the ZCCS code.\\
Fan \textit{et al}. in \cite{fan2007z} proposed ZCCSs, refers to a set of matrices or codes
that exhibits zero correlation zone (ZCZ), for  MC-CDMA. ZCCS's correlation properties allow for the deployment of interference-free MC-CDMA in quasi-synchronous channels without the need for power regulation \cite{chen2007next}. For an $(M,N,L,Z)$-ZCCS, we have $M\leq N\lfloor\frac{L}{Z}\rfloor$, where $M,N,L,Z$ refers to the set size, number of sub-carrier, length and ZCZ width respectively. When the equality sign is true, ZCCS is said to be an optimal \cite{liu2011correlation}. Binary sequences are easy to use electronically because of modulo-$2$ arithmetic. Furthermore, modulo-$2$ arithmetic is isomorphic with the use of $\{1,-1\}$, which simplifies both the modulation and correlation processes in signal processing \cite{proakis2007digital}. However, it is often challenging to obtain flexible lengths for binary sequences.\\
In the literature, numerous ZCCSs have been proposed by using direct \cite{sarkar2018optimal,wu2018optimal,sarkar2020direct,wu2020z,sarkar2021pseudo,shen2022new,ghosh2022direct,sarkar2020construction,xie2021constructions} and indirect \cite{adhikary2019new,yu2022new,das2020new}   constructions. In \cite{sarkar2018optimal,wu2018optimal,sarkar2020direct,wu2020z} binary ZCCS of length power-of-two have been presented.  In \cite{sarkar2021pseudo,shen2022new,ghosh2022direct} ZCCS of non-power-of-two (NPT) length have been presented through generalized Boolean function (GBF); however, they are unable to achieve the binary phase.
Sarkar \textit{et al}. in \cite{sarkar2020construction} proposed $q$-ary  $(2^{n+1},2^{n+1},2^{m-1}+2,2^{m-2}+2^{\psi(m-3)+1})$-ZCCS where $m\geq3$   
using GBF.
In \cite{xie2021constructions}, Sun \textit{et al}. proposed $q$-ary $(2^{k+1},2^{k+1},3.2^{m},2^{m+1})$-ZCCS where $m,k\geq 1$  using GBF. 
Adhikary \textit{et al}. in \cite{b} proposed an optimal binary $(2^{n+1},2^{n+1},N,\frac{N(N+1)}{2})$-ZCCS by indirect method, where $N=2^{\alpha+1}10^{\beta}26^{\gamma},\alpha,\beta~\text{and}~ \gamma\geq 1.$
\\
Motivated by the scarcity of NPT-length binary optimal ZCCS,
this letter proposes three GBF-based direct constructions of binary optimal ZCCS with length $R\gamma, R2^{m_{2}}$ and $3\gamma$ where $\gamma=2^{m_{1}-1}+2^{m_{1}-3}, m_{1}\geq 5,m_{2}\geq 1$ and $R$ is any even number. Furthermore, for the first time, the suggested design provides an NPT length set size of the kind $R2^{k+1}$ for
$k\geq 1$. 
 So, the suggested constructions make a new class of ZCCSs with a length and set size that has never been reported before. \\ 
 
\label{sec:intro}
\vspace{-0.8cm}
\section{Preliminary}
\vspace{-0.2cm}
In this section, we introduce some basic ideas and lemmas that will be applied to the rest of the proposed building.
\par Let 
$\textbf{u}=(u_{0},\hdots,u_{L-1})$ and $\textbf{v}=(v_{0},\hdots ,v_{L-1})$ consist of two complex valued  sequences. 
We define the aperiodic cross-correlation sum (ACCS) between $\textbf{u}$ and $\textbf{v}$ as
\begin{equation}\label{equ:cross}
\Theta(\textbf{u}, \textbf{v})({\tau})=\begin{cases}
\sum_{k=0}^{L-1-\tau}u_{k}v^{*}_{k+\tau}, & 0 \leq \tau < L, \\
\sum_{k=0}^{L+\tau -1}u_{k-\tau}v^{*}_{k}, & -L< \tau < 0,  \\
0, & \text{otherwise},
\end{cases}
\end{equation}
where $*$ stands for the complex conjugate. If $\textbf{u}=\textbf{v}$, then the corresponding function is known as the aperiodic auto-correlation sum (AACS) of $\textbf{u}$, refer to $\psi(\textbf{u})$.
 \par Let $\mathbf{C}_{i}=\{\textbf{u}_{k}^{i}:0\leq k<N\}$ and $\mathbf{C}_{j}=\{\textbf{u}_{k}^{j}:0\leq k<N\}$  be  collections of $N$ sequences, where $\textbf{u}_{k}^{i}=({u}_{k,0}^{i},\ldots,{u}_{k,L-1}^{i})$ and $\textbf{u}_{k}^{j}=({u}_{k,0}^{j},\ldots,{u}_{k,L-1}^{j})$. For two sets of sequence $\mathbf{C}_{i}$ and $\mathbf{C}_{j}$, the ACCS is defined by
 \vspace{-0.3cm}
 $$\Theta(\textbf{C}_{i},\textbf{C}_{j})({\tau})=\sum_{k=0}^{N-1}\Theta(\textbf{u}_{k}^{i},\textbf{u}_{k}^{j}).$$

\vspace{-0.3cm}
\begin{definition}
Let $\mathbf{C}=\{\mathbf{C}_{0},\ldots,\mathbf{C}_{M-1}\}$ be collection of $M$ such sequence sets. The code set $\mathbf{C}$ is called a
$(M,N,L,Z)$-ZCCS (\cite{sarkar2018optimal}) if
\begin{eqnarray}
\Theta(\mathbf{C}_{\mu_1},\mathbf{C}_{\mu_2})(\tau)
=\begin{cases}
NL, & \tau=0, \mu_1=\mu_2,\\
0, & 0<|\tau|<Z, \mu_1=\mu_2,\\
0, & |\tau|< Z, \mu_1\neq \mu_2,
\end{cases}
\end{eqnarray}
\end{definition}
where $0\leq \mu_{1},\mu_{2}\leq M-1$. When $M=N$ and $Z=L$, we denote the set  $\mathbf{C}$ by $(M,M,L)$-CCC.
\vspace{-0.5cm}
\subsection{Generalized Boolean Functions (GBFs)}
A generalized Boolean function (GBF) $f$ with $m$ variables is a mapping from the set $\{0,1\}^m$ to the set $\mathbb{Z}_q=\{0,\ldots,q-1\}$ where, $q$ is a positive integer. The sequence of a GBF $f$ is a $\mathbb{Z}_{q}$-valued sequence of length $N=2^{m}$ defined by $\psi(f)=(\omega_{q}^{f_0},\hdots, \omega_{q}^{f_{2^m-1}})$ where $\omega_{q}=exp(\frac{2\pi \sqrt{-1}}{q}), f_{r}=f(r_0,\hdots,r_{m-1})$ and $(r_0,\hdots,r_{m-1})$ is the binary representation of the integer $r$. We denote $\psi_{j}(f)$ be the truncated sequence after taking first $j$ terms from $\psi(f)$ i.e., $\psi_{j}(f)=(\omega_{q}^{f_{0}},\ldots,\omega_{q}^{f_{j-1}})$ and $\psi^{j}(f)$  be the truncated sequence after taking last $j$ terms from $\psi(f)$ i.e, $\psi^{j}(f)=(\omega_{q}^{f_{2^{m}-j}},\ldots,\omega_{q}^{f_{2^{m}-1}})$.  Let $\mathcal{C}=\{f^{1},\ldots,f^{r}\}$ represents an ordered set of $r$ GBFs, each with $m$ variables. We define the codes $\psi(\mathcal{C})=(\psi(f^{1}),\ldots,\psi(f^{r}))$, $\psi_{j}(\mathcal{C})=(\psi_{j}(f^{1}),\ldots,\psi_{j}(f^{r}))$ and $\psi^{j}(\mathcal{C})=(\psi^{j}(f^{1}),\ldots,\psi^{j}(f^{r}))$. 
\begin{lemma}(Construction of CCC\cite{kumar2023direct})\label{lemma3}\\
Let $m_{1}\geq 5$, $q=2$, $Q$ be a quadratic form  in $m_{1}-4$ variables. Let $G(Q)$ be the labeled graph corresponding to $Q$ (see \cite{kumar2023direct}).  Assume  that $G(Q)$ has  $k\leq m_{1}-5$  distinct vertices labeled by $0\leq p_{0}<p_{1}<\cdots <p_{k-1}<m-4$ such that after removing those $k$ vertices along with all of the edges connecting them yields a path. Let $\beta_{1}$ be any of the end vertices of the path.
For any $d,d_i \in \mathbb{Z}_2$, let us define a GBF $g:\{0,1\}^{m_{1}}\rightarrow \mathbb{Z}_2$ as
\vspace{-0.4cm}
$$
g(z_{0},\ldots,z_{m_{1}-1})=Q+\sum_{i=0}^{m_{1}-5}d_iz_i+d+\alpha+\beta,
$$
\vspace{-0.05cm}
where $\alpha=\Bar z_{m_{1}-1}\left(\Bar z_{m_{1}-4}\left(z_{m_{1}-3}+z_{m_{1}-2}\right)+z_{m_{1}-2}z_{m_{1}-3}\right),$ $\beta=z_{\beta_1}(\Bar z_{m_{1}-1}\left(z_{m_{1}-2}\Bar z_{m_{1}-3}\Bar z_{m_{1}-4}+z_{m_{1}-2}z_{m_{1}-3}\right)+z_{m_{1}-1}\Bar z_{m_{1}-2}\Bar z_{m_{1}-3})$.
Let us consider the functions $g^{\mathbf{a},n}:\{0,1\}^{m_{1}}\rightarrow\mathbb{Z}_{2}$ and $s^{\mathbf{a},n}:\{0,1\}^{m_{1}}\rightarrow\mathbb{Z}_{2}$ as
\begin{eqnarray}
\label{Hari}
g^{\mathbf{a},n}(z_{0},\ldots,z_{m_{1}-1})=g+\sum_{\alpha=0}^{k-1} \left(a_{\alpha}+ n_{\alpha}\right)z_{p_{\alpha}}+a z_{\beta_{1}},\\
\label{bol}
s^{\mathbf{a},n}(z_{0},\ldots,z_{m_{1}-1})=\Bar{g}+\sum_{\alpha=0}^{k-1} \left(a_{\alpha} +n_{\alpha}\right)\Bar z_{p_{\alpha}}+\Bar{a} z_{\beta_{1}},
\end{eqnarray}
where $\Bar{g}(z_{0},\ldots,z_{m_{1}-1})=  g(\Bar{z}_{0},\ldots,\Bar{z}_{m_{1}-1}),\Bar z_{p_{\alpha}}=1- z_{p_{\alpha}},$\\$\Bar{z_{i}}=1-z_{i},\Bar{a}=1-a$, $a\in\{0,1\}$,  $\mathbf{a}=(a_{0},\ldots,a_{k-1},a)\in$\\$\{0,1\}^{k+1}$, $(n_{0},\ldots,n_{k-1})$ is the binary vector representation of the integer $n$ for $0\leq n<2^k$. 
Define an ordered set of GBF $ S_n= \left\{g^{\mathbf{a},n}: \mathbf{a} \in \{0,1\}^{k+1}\right\}$ and $\Bar{S}_n=\left\{s^{\mathbf{a},n}: \mathbf{a} \in \{0,1\}^{k+1}\right\}.$
Then the code set 
$\left\{\psi_{\gamma}(S_n): 0\leq n<2^k\right\}$$
    \cup\left\{\left(\psi^{\gamma}(\Bar S_n)\right)^{*}: 0\leq n<2^k\right\}$ forms a binary $(2^{k+1},2^{k+1},2^{m_{1}-1}+2^{m_{1}-3})$-CCC where $\left(\psi^{\gamma}(\Bar S_n)\right)^{*}$ is the complex conjugate of $\psi^{\gamma}(\Bar S_n)$ and $\gamma=2^{m_{1}-1}+2^{m_{1}-3}$. 
\end{lemma}
   Let $g^{\mathbf{a},n}$ and $s^{\mathbf{a},n}$ be the GBFs of $m_{1}$ variables $z_{0},\ldots,$\\$z_{m_{1}-1}$ as defined in (\ref{Hari}) and (\ref{bol}) respectively. Consider a set
   \begin{equation}
       S=\{\mathbf{c}=({c}_0,\hdots, {c}_{l-1}):0\leq c_{i}\leq 1\}.
   \end{equation}
   We define the GBFs $M^{\mathbf{a},n,\mathbf{c}}:\{0,1\}^{m_{1}+l}\rightarrow \mathbb{Z}_{2}$ and $N^{\mathbf{a},n,\mathbf{c}}:\{0,1\}^{m_{1}+l}\rightarrow \mathbb{Z}_{2}$ with the help of $g^{\mathbf{a},n}$ and $s^{\mathbf{a},n}$ as $M^{\mathbf{a},n,\mathbf{c}}=g^{\mathbf{a},n}+\sum_{i=0}^{l-1}{c}_{i}z_{m_{1}+i}$ and $N^{\mathbf{a},n,\mathbf{c}}=s^{\mathbf{a},n}+\sum_{i=0}^{l-1}\!{c}_{i}z_{m_{1}+i}$ respectively.

The sequence $\psi(M^{\mathbf{a},{n},\mathbf{c}})$ and $\psi(N^{\mathbf{a},{n},\mathbf{c}})$ can be viewed as $2^l$ parts, i.e., $\psi(M^{\mathbf{a},{n},\mathbf{c}})=(\mathbf{u}_{0},\ldots,\mathbf{u}_{2^{l}-1}),~\psi(N^{\mathbf{a},{n},\mathbf{c}})=(\mathbf{v}_{0},\ldots,\mathbf{v}_{2^{l}-1}),$
where $\mathbf{u}_{j}=\psi(g^{\mathbf{a},{n}})\omega_{2}^{\sum_{i=0}^{l-1}c_{i}j_{i}}$,  $\mathbf{v}_{j}=\psi(s^{\mathbf{a},{n}})\omega_{2}^{\sum_{i=0}^{l-1}c_{i}j_{i}}$, $(j_{0},\ldots,j_{l-1})$ is the binary representation of $j$ and $0\leq j\leq 2^{l}-1$. 
Let $\gamma=2^{m_{1}-1}+2^{m_{1}-3}$, we define a truncated sequence $\mathcal{M}^{\mathbf{a},{n},\mathbf{c}}_{\gamma,R}=(\mathbf{u}^{\gamma}_{0},\ldots,\mathbf{u}^{\gamma}_{R-1})$ and $\mathcal{N}^{\mathbf{a},{n},\mathbf{c}}_{\gamma,R}=(\mathbf{v}^{\gamma}_{0},\ldots,\mathbf{v}^{\gamma}_{R-1})$ obtained from  $\psi(M^{\mathbf{a},{n},\mathbf{c}})$ and $\psi(N^{\mathbf{a},{n},\mathbf{c}})$ respectively where  $\mathbf{u}^{\gamma}_{r}=\psi_{\gamma}(g^{\mathbf{a},{n}})\omega_{2}^{\sum_{i=0}^{l-1}c_{i}r_{i}}$, $\mathbf{v}^{\gamma}_{r}=\psi^{\gamma}(s^{\mathbf{a},\mathbf{n}})\omega_{2}^{\sum_{i=0}^{l-1}c_{i}r_{i}}$, $0\leq r< R$, $R$ is an even number such that $2\leq R\leq 2^{l}$, and $(r_{0},\ldots,r_{l-1})$ is the binary vector representation of the integer $r$. Now, we define \begin{eqnarray}
 \label{7401}
 \Omega_n^\mathbf{c}\!\!=\!\!\Big\{\mathcal{M}_{\gamma,R}^{\mathbf{a},{n},\mathbf{c}}:\mathbf{a}\in \{0,1\}^{k+1}\Big\},\\
 \label{7412}
 \Lambda_n^\mathbf{c}\!\!=\!\!\Big\{\left(\mathcal{N}_{\gamma,R}^{\mathbf{a},{n},\mathbf{c}}\right)^{*}:\mathbf{a}\in \{0,1\}^{k+1}\Big\}.
 \end{eqnarray}
 \vspace{-0.5cm}
 \section{First Construction of ZCCS} 
 In this section, we provide the construction of optimal binary  $\left(R2^{k},2^{k+1}, R\gamma,\gamma\right)$-ZCCS, where $\gamma=2^{m_{1}-1}+2^{m_{1}-3}$.
\begin{theorem}
 Let  $m_{1}\geq 5$, $q=2$ , $S_{R}\subseteq S$ containing any $R$ elements from the set $S$ and suppose $\Omega_n^\mathbf{c}$ and $\Lambda_n^\mathbf{c}$ be the set of sequences defined in (\ref{7401}) and (\ref{7412}).
 Then the code set\\
 $\mathcal{G}=\bigg\{\Omega_{n}^{\mathbf{c}},\Lambda_{n}^{\mathbf{c}}:~\!\!0\!\!\leq\!\! n\!\!<\!2^{k},\mathbf{c}\in S_R\bigg\}$
forms an optimal binary  $(R2^{k+1},2^{k+1},R\gamma,\gamma)$-ZCCS over $\mathbb{Z}_{2}$, where  $R\geq 2$ is an even integer and $R\leq2^{l}$.
\end{theorem}
\begin{IEEEproof}
 Let $(\alpha_{0},\ldots,\alpha_{l-1})$ be the vector representation of an integer $\alpha$, where $0\leq\alpha\leq R-1$. For $\mathbf{c},\mathbf{c}'\in S_R,0\leq n,n'<2^{k},\tau=0$, we have
\begin{equation}
\label{1312}
    \begin{split}
        &\Theta\left(\Omega_n^\mathbf{c},\Omega_{n'}^{\mathbf{c} '}\right)(0)\\
        &=\displaystyle\sum_{\mathbf{a}}\Theta\left(\mathcal{M}_{\gamma,R}^{\mathbf{a},{n},\mathbf{c}},\mathcal{M}_{\gamma,R}^{\mathbf{a},{n}',\mathbf{c}'}\right)(0)\\
        &=\displaystyle\sum_{\mathbf{a}}\Theta\left({\psi_{\gamma}}(g^{\mathbf{a},\mathbf{n}}),{\psi_{\gamma}}(g^{\mathbf{a},\mathbf{n}'})\right)(0)
        \sum_{\alpha=0}^{R-1}\prod_{i=0}^{l-1}\omega_{2}^{(c_i-c_i')\alpha_{i}}\\
        &=\Theta\Big(\psi_{\gamma}(S_n),\psi_{\gamma}(S_{n'})\Big)(0)\sum_{\alpha=0}^{R-1}\prod_{i=0}^{l-1}\omega_{2}^{(c_i-c_i')\alpha_{i}}\\
        &=\begin{cases}
     R2^{k+1}\gamma,& (n,\mathbf{c})=(n',\mathbf{c}'),\\
     0         ,& (n,\mathbf{c})\neq(n',\mathbf{c}').
    \end{cases}
    \end{split}
\end{equation}
Now, by (\ref{7401}), (\ref{7412}) and \textit{Lemma} \textit{1},  the ACCS between $\Omega_n^\mathbf{c}$ 
and $\Omega_{n'}^{\mathbf{c}^{'}}$ for $0<|\tau|<\gamma$ can be expressed as
\begin{equation}
\label{14}
    \begin{split}
        &\Theta\left(\Omega_n^\mathbf{c},\Omega_{n^{'}}^{\mathbf{c}^{'}}\right)(\tau)\\
        &=\displaystyle\sum_{\mathbf{a}}\Theta\left({\psi_{\gamma}}(g^{\mathbf{a},\mathbf{n}}),{\psi_{\gamma}}(g^{\mathbf{a},\mathbf{n}'})\right)(\tau)\sum_{\alpha=0}^{R-1}\prod_{i=0}^{l-1}\omega_{2}^{(c_i-c_i')\alpha_{i}}\\
        &+\!\!\displaystyle\sum_{\mathbf{a}}\!\!\Theta\!\!\left({\psi_{\gamma}}(g^{\mathbf{a},\mathbf{n}}),{\psi_{\gamma}}(g^{\mathbf{a},\mathbf{n}'})\!\!\right)\!\!(\tau-\gamma)\sum_{\alpha=0}^{R-2}\prod_{i=0}^{l-1}\omega_{2}^{c_{i}(\alpha_{i})-c'_{i}(\alpha'_{i})}\\
        &=\Theta\Big(\psi_{\gamma}(S_{n}),\psi_{\gamma}(S_{n^{'}})\Big)(\tau)\sum_{\alpha=0}^{R-1}\prod_{i=0}^{l-1}\omega_{2}^{(c_i-c_i')\alpha_{i}}\\
         &+\Theta\Big(\psi_{\gamma}(S_{n}),\psi_{\gamma}(S_{n^{'}})\Big)(\tau-\gamma)\sum_{\alpha=0}^{R-2}\prod_{i=0}^{l-1}\omega_{2}^{c_{i}(\alpha_{i})-c'_{i}(\alpha'_{i})},
    \end{split}
\end{equation}
\!where $(\alpha'_{0},\ldots,\alpha'_{l-1})$ is the binary representation of $\alpha+1$. From \textit{Lemma} \textit{1}, we obtain
 \begin{equation}
 \label{1912}
    \Theta\Big(\psi_{\gamma}(S_{n}),\psi_{\gamma}(S_{n^{'}})\Big)(\tau)=0,0< |\tau|<\gamma.
 \end{equation}
 From (\ref{14}) and (\ref{1912}), we have
 \vspace{-0.2cm}
\begin{equation}
\label{15}
 \begin{split}
    \Theta\Big(\Omega_n^\mathbf{c},\Omega_{n^{'}}^{\mathbf{c} '}\Big)(\tau)=0, 0<|\tau|<\gamma.
 \end{split}
\end{equation}
In the similar way, it can be shown that
\begin{equation}\label{17}
 \begin{split}
  \Theta&\Big(\Lambda_n^\mathbf{c},\Lambda_{n'}^{\mathbf{c}'}\Big)(\tau)=\begin{cases}
     R2^{k+1}\gamma,& 
     (n,\mathbf{c})=(n',\mathbf{c}'),\tau=0,\\
     0         ,& \begin{aligned}
         &(n,\mathbf{c})\neq(n',\mathbf{c}'),0\leq|\tau|<\gamma.
     \end{aligned}
    \end{cases}
 \end{split}
\end{equation}
Now, by (\ref{7401}), (\ref{7412}) and \textit{Lemma} \textit{1},  the ACCS between $\Omega_n^\mathbf{c}$ 
and $\Lambda_{n'}^{\mathbf{c}^{'}}$ for $\tau=0$ can be expressed as
\begin{equation}
\label{1913051}
    \begin{split}
        &\Theta\left(\Omega_n^\mathbf{c},\Lambda_{n'}^{\mathbf{c} '}\right)(0)\\
        &=\Theta\Big(\psi_{\gamma}(S_n),\left(\psi^{\gamma}(\Bar{S}_{n'})\right)^{*}\Big)(0)\sum_{\alpha=0}^{R-1}\prod_{i=0}^{l-1}\omega_{2}^{(c_i+c_i')\alpha_{i}}
    \end{split}
\end{equation}
From \textit{Lemma} \textit{1}, we obtain
 \vspace{-0.2cm}
 \begin{equation}
 \label{191306}
    \Theta\Big(\psi_{\gamma}(S_{n}),\left(\psi^{\gamma}(\Bar{S}_{n^{'}})\right)^{*}\Big)(0)=0.
 \end{equation}
From (\ref{1913051}) and (\ref{191306}) we have 
\begin{equation}
\label{191305}
    \begin{split}
        &\Theta\left(\Omega_n^\mathbf{c},\Lambda_{n'}^{\mathbf{c} '}\right)(0)=0.
    \end{split}
\end{equation}
Now, by (\ref{7401}), (\ref{7412}) and \textit{Lemma} \textit{1},  the ACCS between $\Omega_n^\mathbf{c}$ 
and $\Lambda_{n'}^{\mathbf{c}^{'}}$ for $0<|\tau|<\gamma$ can be expressed as
\begin{equation}
\label{191307}
    \begin{split}
&\Theta\left(\Omega_n^\mathbf{c},\Lambda_{n^{'}}^{\mathbf{c}^{'}}\right)(\tau)\\
        &=\Theta\Big(\psi_{\gamma}(S_{n}),\left(\psi^{\gamma}(\Bar{S}_{n^{'}})\right)^{*}\Big)(\tau)\sum_{\alpha=0}^{R-1}\prod_{i=0}^{l-1}\omega_{2}^{(c_i+c_i')\alpha_{i}}\\
         &+\Theta\Big(\psi_{\gamma}(S_{n}),\left(\psi^{\gamma}(\Bar{S}_{n^{'}})\right)^{*}\Big)(\tau-\gamma)\sum_{\alpha=0}^{R-2}\prod_{i=0}^{l-1}\omega_{2}^{c_{i}(\alpha_{i})+c'_{i}(\alpha'_{i})}.
    \end{split}
\end{equation}
From \textit{Lemma} \textit{1}, we obtain
 \vspace{-0.2cm}
 \begin{equation}
 \label{191309}
    \Theta\Big(\psi_{\gamma}(S_{n}),\left(\psi^{\gamma}(\Bar{S}_{n^{'}})\right)^{*}\Big)(\tau)=0,0<|\tau|<\gamma.
 \end{equation}
From (\ref{191307}) and (\ref{191309}) we have 
\begin{equation}
\label{1913010}
    \begin{split}
        &\Theta\left(\Omega_n^\mathbf{c},\Lambda_{n'}^{\mathbf{c} '}\right)(\tau)=0,0<|\tau|<\gamma.
    \end{split}
\end{equation}

\vspace{-0.1cm}
Thus, from (\ref{1312}), (\ref{15}), (\ref{17}), (\ref{191305}), and (\ref{1913010}),   it can be concluded that the set $\mathcal{G}$
 forms an optimal binary  $\left(R2^{k+1},2^{k+1},R\gamma,\gamma\right)$-ZCCS.
\end{IEEEproof}
\vspace{-0.45cm}
\section{Second Construction of ZCCS}
In this section we provide optimal binary  $(R2^{k+1},2^{k+1},$\\$R2^{m_{2}},2^{m_{2}})$-ZCCS through GBF.
\begin{lemma}(Construction of CCC\cite{rathinakumar2008complete})\label{lemma3}\\
Let $q$ be an even, $m_{2}\geq 1$, $f$ be a GBF of $m_{2}$ variables. Consider a graph $G(f)$ which contains a set of $k$ unique vertices labeled as $0\leq p_0<\cdots<p_{k-1}<m_{2}$ such that if these $k$ vertices  are removed from the graph, it will be a path. Let $\beta_{1}$ be any of the end vertices of the path.
Let us consider the functions $f^{\mathbf{a},n}:\{0,1\}^{m_{2}}\rightarrow\mathbb{Z}_{q}$ and $h^{\mathbf{a},n}:\{0,1\}^{m_{2}}\rightarrow\mathbb{Z}_{q}$ as $f^{\mathbf{a},n}(z_{0},\ldots,z_{m_{2}-1})\!\!=\!\!{f}+
 \frac{q}{2}\Big(\sum_{\alpha=0}^{k-1} \left(a_{\alpha}+n_{\alpha}\right)z_{p_{\alpha}}+a z_{\beta_{1}}\Big)$, $h^{\mathbf{a},n}(z_{0},\ldots,z_{m_{2}-1})\!\!=\Bar{f}+
 \frac{q}{2}\Big(\sum_{\alpha=0}^{k-1} \!\!\left(a_{\alpha}\!+\!n_{\alpha}\right)\!\Bar{z}_{p_{\alpha}}\!\!+\Bar{a} z_{\beta_{1}}\Big).$
Define an ordered sets $ G_n= \left\{f^{\mathbf{a},n}: \mathbf{a} \in \{0,1\}^{k+1}\right\}$ and $\Bar G_n=\left\{h^{\mathbf{a},n}: \mathbf{a} \in \{0,1\}^{k+1}\right\}.$
 Then
  $\{\psi(G_n),\left(\psi(\Bar{G}_{n})\right)^{*}:0\leq n<2^k\}$ forms $q$-ary
$(2^{k+1},2^{k+1},2^{m_{2}})$-CCC.
\end{lemma}
We define a GBF $R^{\mathbf{a},n,\mathbf{c}}:\{0,1\}^{m_{2}+l}\rightarrow \mathbb{Z}_{q}$ and $S^{\mathbf{a},n,\mathbf{c}}:\{0,1\}^{m_{2}+l}\rightarrow \mathbb{Z}_{q}$ as $R^{\mathbf{a},n,\mathbf{c}}=f^{\mathbf{a},n}+\frac{q}{2}\sum_{i=0}^{l-1}{c}_{i}z_{m_{2}+i}$ and $S^{\mathbf{a},n,\mathbf{c}}=h^{\mathbf{a},n}+\frac{q}{2}\sum_{i=0}^{l-1}\!{c}_{i}z_{m_{2}+i}$ respectively. 
Therefore, we have $\psi(R^{\mathbf{a},{n},\mathbf{c}})=(\mathbf{e}_{0},\ldots,\mathbf{e}_{2^{l}-1}),\psi(S^{\mathbf{a},{n},\mathbf{c}})=(\mathbf{p}_{0},\ldots,$\\$\mathbf{p}_{2^{l}-1}),$
where $\mathbf{e}_{j}=\psi(f^{\mathbf{a},{n}})\omega_{2}^{\sum_{i=0}^{l-1}c_{i}j_{i}}$, $\mathbf{p}_{j}=\psi(h^{\mathbf{a},{n}})\omega_{2}^{\sum_{i=0}^{l-1}c_{i}j_{i}}$. 
We define truncated sequence $\mathcal{U}^{\mathbf{a},{n},\mathbf{c}}=(\mathbf{e}_{0},\ldots,\mathbf{e}_{R-1})$ and $\mathcal{V}^{\mathbf{a},{n},\mathbf{c}}= (\mathbf{p}_{0},\ldots,\mathbf{p}_{R-1})$ obtained from $\psi(R^{\mathbf{a},{n},\mathbf{c}})$ and $\psi(S^{\mathbf{a},{n},\mathbf{c}})$ respectively.
~Now, in the similar way as \textit{Lemma} 1  we can define \begin{eqnarray}
 \label{740}
 \Phi_n^\mathbf{c}\!\!=\!\!\Big\{\mathcal{U}^{\mathbf{a},{n},\mathbf{c}}:\mathbf{a}\in \{0,1\}^{k+1}\Big\},\\
 \label{741}
 \Delta_n^\mathbf{c}\!\!=\!\!\Big\{\left(\mathcal{V}^{\mathbf{a},{n},\mathbf{c}}\right)^{*}:\mathbf{a}\in \{0,1\}^{k+1}\Big\}.
 \end{eqnarray}
 \begin{theorem}
Consider a set $S=\{\mathbf{c}=({c}_1,\hdots, {c}_l):c_{i}\in\{0,1\}\}$, suppose $\Phi_n^\mathbf{c}$ and $\Delta_n^\mathbf{c}$ be the set of sequences as defined in (\ref{740}) and (\ref{741}) respectively. Let $S_{R}\subseteq S$ consists any $R$ elements from the set $S$,
 then the code set
$\mathcal{G}=\bigg\{\Phi_{n}^{\mathbf{c}},\Delta_{n}^{\mathbf{c}}:~\!\!0\!\!\leq\!\! n\!\!<\!2^{k},\mathbf{c}\in S_R\bigg\}$
forms an $q$-ary  $(R2^{k},2^{k+1},R2^{m_{2}},2^{m_{2}})$-ZCCS over $\mathbb{Z}_{2}$, where $q,R\geq 2$ are even integers and $R\leq2^{l}$.
\end{theorem}
\begin{IEEEproof}
    The proof is exactly analogous to the \textit{Theorem} 1.
\end{IEEEproof}
\vspace{-0.4cm}
\section{Third Construction of ZCCS}
In this section, we provide optimal binary $(2^{k+1},2^{k+1},$\\$3\gamma,2\gamma)$-ZCCS construction by using GBF.\\
We define the function $F^{\mathbf{a},n}:\{0,1\}^{m_{1}+2}\rightarrow \mathbb{Z}_{2}$ and  $G^{\mathbf{a},n}:\{0,1\}^{m_{1}+2}\rightarrow \mathbb{Z}_{2}$ with the help of  $g^{\mathbf{a},n}$ and $s^{\mathbf{a},n}$ as $F^{\mathbf{a},n}=g^{\mathbf{a},n}+z_{m_{1}+1}$ and $G^{\mathbf{a},n}=s^{\mathbf{a},n}+z_{m_{1}+1}$ respectively.
The sequences $\psi(F^{\mathbf{a},n})$ and $\psi(G^{\mathbf{a},n})$ can be written as
\begin{equation}
    \begin{split}
      &\psi(F^{\mathbf{a},n})=\big(\psi(g^{\mathbf{a},n}),\psi(g^{\mathbf{a},n}),-\psi(g^{\mathbf{a},n}),-\psi(g^{\mathbf{a},n})\big),\\
     &\psi(G^{\mathbf{a},n})=\big(\psi(s^{\mathbf{a},n}),\psi(s^{\mathbf{a},n}),-\psi(s^{\mathbf{a},n}),-\psi(s^{\mathbf{a},n})\big).\\
    \end{split}
\end{equation}
We define the truncated  sequence $\psi_{T}(F^{\mathbf{a},n})$ and $\psi_{T}(G^{\mathbf{a},n})$ obtained from $\psi(F^{\mathbf{a},n})$ and 
$\psi(G^{\mathbf{a},n})$ as 
\begin{eqnarray}
\label{1654}
\psi_{T}(F^{\mathbf{a},n})=\big(\psi_{\gamma}(g^{\mathbf{a},n}),\psi_{\gamma}(g^{\mathbf{a},n}),-\psi_{\gamma}(g^{\mathbf{a},n})\big),\\
\label{1655}
\psi_{T}(G^{\mathbf{a},n})=\big(\psi^{\gamma}(s^{\mathbf{a},n}),\psi^{\gamma}(s^{\mathbf{a},n}),-\psi^{\gamma}(s^{\mathbf{a},n})\big).
\end{eqnarray}
where $\psi_{\gamma}(g^{\mathbf{a},n})=(\omega_{2}^{g_{0}^{\mathbf{a},n}},\ldots,\omega_{2}^{g_{\gamma-1}^{\mathbf{a},n}})$ and $\psi^{\gamma}(s^{\mathbf{a},n})=(\omega_{2}^{s_{2^{m}-\gamma}^{\mathbf{a},n}},\ldots,\omega_{2}^{s_{2^{m}-1}^{\mathbf{a},n}})$, $g_{r}^{\mathbf{a},n}=g^{\mathbf{a},n}(r_{0},\ldots,r_{m-1})$, $s_{r}^{\mathbf{a},n}=s^{\mathbf{a},n}(r_{0},\ldots,r_{m-1})$, $(r_{0},\ldots.r_{m-1})$ is the binary vector representation of the integer $r$ and $0\leq r\leq 2^{m}-1$  .
\begin{theorem}
Let $\psi_{T}(F^{\mathbf{a},n})$ and $\psi_{T}(G^{\mathbf{a},n})$ be the sequences as defined in (\ref{1654}) and (\ref{1655}) respectively. We define the order sets $\mathbf{C}_n=\{\psi_{T}(F^{\mathbf{a},n}):\mathbf{a}\in\{0,1\}^{k+1}\},$ $\Bar{\mathbf{C}}_n=\{\psi_{T}(G^{\mathbf{a},n}):\mathbf{a}\in\{0,1\}^{k+1}\}.$ Then the code set $\left\{\mathbf{C}_n: 0\leq n<2^k\right\}
    \cup\left\{\left(\Bar{\mathbf{C}}_n\right)^{*}: 0\leq n<2^k\right\},$
forms an optimal binary $(2^{k+1},2^{k+1},3\gamma,2\gamma)$-ZCCS.
\end{theorem}

\begin{IEEEproof}
For $0\leq n,n'<2^{k}$, $\tau=0$ the AACS between $\mathbf{C}_{n}$ and $\mathbf{C}_{n'}$ can be derive as
\begin{equation}
\label{16111}
    \begin{split}
      \Theta\Big(\mathbf{C}_{n},\mathbf{C}_{n'}\Big)(0)
      &=\displaystyle\sum_{\mathbf{a}}\Theta\Big(\psi_{T}(F^{\mathbf{a},{n}}),\psi_{T}(F^{\mathbf{a},{n'}})\Big)(0)\\
      &=3\displaystyle\sum_{\mathbf{a}}\Theta\Big(\psi_{\gamma}(g^{\mathbf{a},{n}}),\psi_{\gamma}(g^{\mathbf{a},{n'}})\Big)(0)\\
      &=3\Theta\Big(\psi_{\gamma}(S_{n}),\psi_{\gamma}(S_{n'})\Big)(0)
    \end{split}
\end{equation}
By using \textit{Lemma 1} and  (\ref{16111}), we have 
\begin{equation}
\label{161220}
    \begin{split}
      \Theta\Big(\mathbf{C}_{n},\mathbf{C}_{n'}\Big)(0)=\begin{cases}
      32^{k+1}\gamma & n=n';\\
      0 & n\neq n'.
      \end{cases}
    \end{split}
\end{equation}
When $0<\tau<\gamma$
\begin{equation}
\label{R}
\begin{split}
     &\Theta\Big(\mathbf{C}_{n},\mathbf{C}_{n'}\Big)(\tau)\\
     &=3\Theta\Big(\psi_{\gamma}(S_{n}),\psi_{\gamma}(S_{n'})\Big)(\tau)\!+\!\Theta\Big(\psi_{\gamma}(S_{n}),\psi_{\gamma}(S_{n'})\Big)(\gamma-\tau)\\
     &-\Theta\Big(\psi_{\gamma}(S_{n}),\psi_{\gamma}(S_{n'})\Big)(\gamma-\tau).
\end{split}
\end{equation}
\vspace{-0.25cm}
Using  \textit{Lemma 1} and (\ref{R}), we have
\begin{equation}
\label{161221}
    \Theta\big(\mathbf{C}_{n},\mathbf{C}_{n'}\big)(\tau)=0.
\end{equation}
\vspace{-0.25cm}
When $\tau=\gamma$
\begin{equation}
\label{1611}
    \begin{split}
      &\Theta\Big(\mathbf{C}_{n},\mathbf{C}_{n'}\Big)(\gamma)\\
      &=\Theta\Big(\psi_{\gamma}(S_{n}),\psi_{\gamma}(S_{n'})\Big)(0)-\Theta\Big(\psi_{\gamma}(S_{n}),\psi_{\gamma}(S_{n'})\Big)(0)\\
      &=0
    \end{split}
\end{equation}
When $\gamma<|\tau|<2\gamma-1$, we have
\vspace{-0.3cm}
\begin{equation}
\label{1917}
    \begin{split}
       &\Theta\Big(\mathbf{C}_{n},\mathbf{C}_{n'}\Big)(\tau)\\ 
       &=\!\!\Theta\Big(\psi_{\gamma}(S_{n}),\psi_{\gamma}(S_{n'})\Big)\!(\tau\!-\!\gamma)\!-\!\Theta\!\Big(\psi_{\gamma}(S_{n}),\!\psi_{\gamma}(S_{n'})\Big)\!(\tau-\gamma)\\
       &-\Theta\Big(\psi_{\gamma}(S_{n}),\psi_{\gamma}(S_{n'})\Big)(2\gamma-\tau).
    \end{split}
\end{equation}
\vspace{-0.2cm}
Using \textit{Lemma 1} and (\ref{1917}), we have
\begin{equation}
\label{161222}
\Theta\Big(\mathbf{C}_{n},\mathbf{C}_{n'}\Big)(\tau)=0.    
\end{equation}
\vspace{-0.2cm}
In a similar way it can be shown that
\begin{equation}
\label{161223}
    \begin{split}
      &\Theta\Big(\!\!\!\left(\Bar{\mathbf{C}}_{n}\right)^{*}\!,\!\left(\Bar{\mathbf{C}}_{n'}\right)^{*}\!\!\Big)\!(\tau)\!=\!\!\begin{cases}
      32^{k+1}\gamma & n=n',\tau=0;\\
      0 & n\neq n',0\leq|\tau|\leq 2\gamma-1.
      \end{cases}
    \end{split}
\end{equation}
\vspace{-0.2cm}
By similar argument,  $\forall\tau,~0\leq|\tau|\leq 2\gamma-1$ and we have
\begin{eqnarray}
\label{161224}
&\Theta\Big(\mathbf{C}_{n},\left(\Bar{\mathbf{C}}_{n'}\right)^{*}\Big)(\tau)=0.
\end{eqnarray}
Hence, from (\ref{161220}),(\ref{161221}),(\ref{161222}),(\ref{161223}) and (\ref{161224}) the result follws.
\end{IEEEproof}
\begin{example}
For $m_{1}=8$, $q=2,l=1, R=2$, let us assume $Q:\{0,1\}^{4} \rightarrow \mathbb{Z}_q$ be a quadratic form in variables $z_0,z_1,z_{2},z_{3}$, i.e., $Q\left(z_0,z_1,z_{2},z_{3}\right)=z_0z_1+z_1z_2+z_2z_3+z_3z_0+z_0z_2.$
Let $d_0=1,d_1=1,d_2=1,d_3=1,d=0$, and define the GBF $g:\{0,1\}^{8}\rightarrow \mathbb{Z}_2$  with the help of $Q$ as $ g(z_{0},\ldots,z_{7})=Q+\sum_{i=0}^{3}d_iz_i+d+\alpha+\beta,$
where $\alpha=\Bar z_7\big(\Bar z_4(z_5+z_6)+z_6z_5\big),\beta=z_2\big(\Bar z_7(z_6\Bar z_5\Bar z_4$ $+z_6z_5)+z_7\Bar z_6\Bar z_5\big).$
 Let $k=2, \mathbf{c}=c_0$ and $\mathbf{a}=(a_{0},a_{1},a)\in\{0,1\}^{3}$ where, $0\leq c_{0}< 1$ and $0\leq n\leq 3$. Consider the function $g^{\mathbf{a},n}:\{0,1\}^{8}\rightarrow\mathbb{Z}_{2}$ and $s^{\mathbf{a},n}:\{0,1\}^{8}\rightarrow\mathbb{Z}_{2}$ as $g^{\mathbf{a},n}=g+\sum_{\alpha=0}^{1} (a_{\alpha}+n_{\alpha}) z_{\alpha}+a z_{1},$ $s^{\mathbf{a},n}=\tilde{g}+\sum_{\alpha=0}^{1} (a_{\alpha}+n_{\alpha})\Bar{z}_{\alpha}+(1-a)z_{1}.$
  Therefore, $M^{\mathbf{a},n,c_{0}}\!\!=g^{\mathbf{a},n}+c_{0}z_{8},$ $N^{\mathbf{a},n,c_{0}}\!\!=s^{\mathbf{a},n}+c_{0}z_{8}.$
 We have $\gamma=160$ and $\mathcal{M}_{160,2}^{\mathbf{a},{n},c_{0}}$ and $\mathcal{N}_{160,2}^{\mathbf{a},{n},c_{0}}$ are the truncated sequences obtained from $M^{\mathbf{a},{n},c_{0}}$ and $N^{\mathbf{a},{n},c_{0}}$ respectively as $\mathcal{M}_{160,2}^{\mathbf{a},{n},c_{0}}=(A^{160}_{0},A^{160}_{1}),$ $\mathcal{N}_{160,2}^{\mathbf{a},{n},c_{0}}=(B^{160}_{0},B^{160}_{1}),$
where, $A^{160}_{0}=\psi_{160}(g^{\mathbf{a},{n}})$, $A^{160}_{1}=\psi_{160}(g^{\mathbf{a},{n}})\omega_{2}^{c_{0}}$, $B^{160}_{0}=\psi^{160}(s^{\mathbf{a},\mathbf{n}})$, $B^{160}_{1}=\psi^{160}(s^{\mathbf{a},\mathbf{n}})\omega_{2}^{c_{0}}.$
Again from (\ref{7401}) and (\ref{7412}), we have $\Omega_n^\mathbf{c}\!\!=\!\!\Big\{\mathcal{M}_{160,2}^{\mathbf{a},{n},\mathbf{c}}:\mathbf{a}\in \{0,1\}^{3}\Big\},$ $\Lambda_n^\mathbf{c}\!\!=\!\!\Big\{\left(\mathcal{N}_{160,2}^{\mathbf{a},{n},\mathbf{c}}\right)^{*}:\mathbf{a}\in \{0,1\}^{3}\Big\}.$
Therefore, by \textit{Theorem 1}, the set $\mathcal{G}=\Big\{\Omega_{n}^{\mathbf{c}},\Lambda_{n}^{\mathbf{c}}:0\leq n\leq 3, 0\leq c_1\leq 1\Big\},$
forms an binary optimal $(16,8,320,160)$-ZCCS.
\end{example}
\begin{example}
For $m_{1}=8$, $q=2$, $k=2$, $\mathbf{a}=(a_{0},a_{1},a)\in\{0,1\}^{3}$ and $0\leq n\leq 3$. Consider the function $g^{\mathbf{a},n}:\{0,1\}^{8}\rightarrow\mathbb{Z}_{2}$ and $s^{\mathbf{a},n}:\{0,1\}^{8}\rightarrow\mathbb{Z}_{2}$ be same as (\ref{Hari}) and (\ref{bol}) respectively. Therefore, we have $F^{\mathbf{a},n}=g^{\mathbf{a},n}+z_{9}$ and $G^{\mathbf{a},n}=s^{\mathbf{a},n}+z_{9}$. After truncation, we left with the sequences $\psi_{T}(F^{\mathbf{a},n})=\big(\psi_{160}(g^{\mathbf{a},n}),\psi_{160}(g^{\mathbf{a},n}),-\psi_{160}(g^{\mathbf{a},n})\big)$, $\psi_{T}(G^{\mathbf{a},n})=\big(\psi^{160}(s^{\mathbf{a},n}),\psi^{160}(s^{\mathbf{a},n}),-\psi^{160}(s^{\mathbf{a},n})\big).$
Therefore, we have $C_n=\{\psi_{T}(F^{\mathbf{a},n}):\mathbf{a}\in\{0,1\}^{3}\},$ $\Bar{C}_n=\{\psi_{T}(G^{\mathbf{a},n}):\mathbf{a}\in\{0,1\}^{3}\}.$
Hence, by \textit{Theorem 3}, we conclude that the set $\left\{C_n: 0\leq n<4\right\}
    \cup\left\{\left(\Bar C_n\right)^{*}: 0\leq n<4\right\},$
forms an optimal binary $(8,8,480,320)$-ZCCS.
\end{example}
\vspace{-0.4cm}
\subsection{Comparative Analysis with Prior Works}
 Using an orthogonal binary ZCP, Adhikary \textit{et al}. in \cite{b} proposed an optimal binary $(2^{n+1},2^{n+1},N,\frac{N(N+1)}{2})$-ZCCS where $N=2^{\alpha+1}10^{\beta}26^{\gamma}$. The GBFs-based constructions in \cite{xie2021constructions,sarkar2020construction} provide binary optimal $(2^{n+1},2^{n+1},2^{m-1}+2,2^{m-2}+2^{\psi(m-3)+1})$-ZCCS and $(2^{k+1},2^{k+1},3.2^{m},2^{m+1})$-ZCCS  respectively. Our approach does not depend on any initial sequence and matrices and provides three new classes of optimal binary $\left(R2^{k+1},2^{k+1}, R\gamma,\gamma\right)$-ZCCS, $\left(R2^{k+1},2^{k+1}, R2^{m_{2}},2^{m_{2}}\right)$-ZCCS and $\left(2^{k+1},2^{k+1},3\gamma,2\gamma\right)$-ZCCS. TABLE I compares our work with  \cite{sarkar2020construction}, \cite{xie2021constructions}, and \cite{b}.
\vspace{-0.5cm}
\begin{table}[h]
\captionsetup{justification=centering, 
 }
\centering
\caption{COMPARISON WITH EXISTING BINARY OPTIMAL NPT ZCCS}
\label{tab:default}
\resizebox{\textwidth}{!}{
\begin{tabular}{|l|l|l|l|l|}
\hline
Source           & Based On       &Parameters  &Conditions      \\ \hline

\cite{sarkar2020construction} &Direct        &$(2^{n+1},2^{n+1},2^{m-1}+2,2^{m-2}+2^{\psi(m-3)+1})$ &$v\leq  m,q\geq2,m\geq2$                             \\ \hline

\cite{xie2021constructions} &Direct        &$(2^{k+1},2^{k+1},3.2^{m},2^{m+1})$ &$m,k\in \mathbb{Z}^+$                              \\ \hline
\cite{b} &{Indirect}         &$(2^{n+1},N,2^{n+1},Z)$   &$N\geq 3$, $N=\text{length of OB-ZCP}$ 
   \\ \hline
\textit{Thm. 1} &Direct          &$\left(R2^{k+1},2^{k+1},R\gamma,\gamma\right)$    &$ m\geq5,k\in\mathbb{Z}^{+},\gamma=2^{m-1}+2^{m-3}$, $R$ is even        
 \\ \hline
\textit{Thm. 2} &Direct          &$\left(R2^{k+1},2^{k+1},R2^{m},2^{m}\right)$    &$ m\geq5,k\in\mathbb{Z}^{+}$, $R$ is even  

\\ \hline
\textit{Thm. 3} &Direct          &$\left(2^{k+1},2^{k+1},3\gamma,2\gamma\right)$    &$ m\geq5,k\in\mathbb{Z}^{+},\gamma=2^{m-1}+2^{m-3}$                          
\\ \hline
\end{tabular}
}
\end{table}

\nocite{pke2015}\nocite{wu2020z}
\begin{remark}
    By the similar calculation as in [9, Remark 3] and [15, Remark 3], \textit{Theorem 1}, and \textit{Theorem 3}, produce at least $2^{m_{1}-3}\frac{(m_{1}-3)!}{2k!}2^{k\choose 2}$ number of non-overlapping ZCCSs and \textit{Theorem 2}, produces at least $q^{m_{2}+1}\frac{m_{2}!}{2k!}(q-1)^{k(m_{2}-k)}q^{\binom{k}{2}}$ number of non-overlapping ZCCSs. 
\end{remark}
\begin{remark}
\cite{sarkar2021pseudo,ghosh2022direct}  offer  NPT-length ZCCS through GBF, but their phase depends on the length. For example, ZCCS of length $60$, \cite{sarkar2021pseudo,ghosh2022direct} have a phase of sequences of $30$, however, we have $2$. Due to noise and interference, transmitted symbols over $Z_{\sigma}$ with large value of $\sigma$ require additional bits for quantization, increasing symbol error probability \cite{lotter1994comparison}.  Our method yields high NPT length ZCCS with binary phase, unlike \cite{sarkar2021pseudo,ghosh2022direct}.

\end{remark}
\vspace{-0.5cm}
\section{Conclusion}
Three families of new optimal binary ZCCSs with novel parameters are proposed in this work. The structures are based on GBFs and we propose three distinct direct methods to generate optimal binary $\left(R2^{k+1},2^{k+1}, R\gamma,\gamma\right)$-ZCCS, $\left(R2^{k+1},2^{k+1}, R2^{m_{2}},2^{m_{2}}\right)$-ZCCS and $\left(2^{k+1},2^{k+1},3\gamma,2\gamma\right)$-ZCCS respectively, where $\gamma=2^{m_{1}-1}+2^{m_{1}-3}, m_{1}\geq 5, k\geq 1,m_{2}\geq 1$ and $R$ is any even number. Our methods create three new classes of binary ZCCSs with unique lengths and set sizes, which are not found in any existing literature. 


\bibliographystyle{IEEEtran}
\bibliography{Bibliography.bib}
\end{document}